\documentclass{svjour3}                     
\smartqed  
\usepackage{graphicx}
\usepackage{bm}
\def\gtap{\ \raise.3ex\hbox{$>$\kern-.75em\lower1ex\hbox{$\sim$}}\ }
\def\ltap{\ \raise.3ex\hbox{$<$\kern-.75em\lower1ex\hbox{$\sim$}}\ }%

\begin{document}

\title{
An $X(3872)$-like peak in $B\to (J/\psi\pi^+\pi^-) K\pi$
due to triangle singularity
\thanks{
This work is in part supported by 
National Natural Science Foundation of China (NSFC) under contracts 
U2032103 and 11625523, and also by
National Key Research and Development Program of China under Contracts 2020YFA0406400.
}
}


\author{Satoshi X. Nakamura}


\institute{S.X. Nakamura \at
University of Science and Technology of China, Hefei 230026, 
People's Republic of China\\
\email{satoshi@ustc.edu.cn}
}

\date{Received: date / Accepted: date}

\maketitle

\begin{abstract}
Triangle mechanisms for 
$B^0\to (J/\psi\pi^+\pi^-) K^+\pi^-$ are studied.
Experimentally, an $X(3872)$ peak has been observed in this process.
When the final $(J/\psi\pi^+\pi^-)\pi$
invariant mass is around the $D^*\bar D^*$ threshold,
one of the triangle mechanisms causes a triangle singularity
and generates a sharp $X(3872)$-like peak in the $J/\psi\pi^+\pi^-$ invariant mass
distribution.
The Breit-Wigner mass and width fitted to the spectrum are 
 3871.68~MeV (a few keV above the $D^{*0}\bar{D}^0$ threshold)
 and $\sim$0.4~MeV, respectively.
These Breit-Wigner parameters hardly depends on a choice of the model parameters.
Comparing with the precisely measured $X(3872)$ mass and width,
$3871.69\pm 0.17$~MeV and $< 1.2$~MeV, 
the agreement is remarkable.
When studying the $X(3872)$ signal from this process, 
this non-resonant contribution has to be understood in advance.
We also study
a charge analogous process $B^0\to (J/\psi\pi^0\pi^-) K^+\pi^0$.
A similar triangle singularity exists and generates an $X^-(3876)$-like peak.
\end{abstract}

\section{Introduction}
\label{sec:intro}
\begin{figure*}[t]
\begin{center}
\includegraphics[width=1\textwidth]{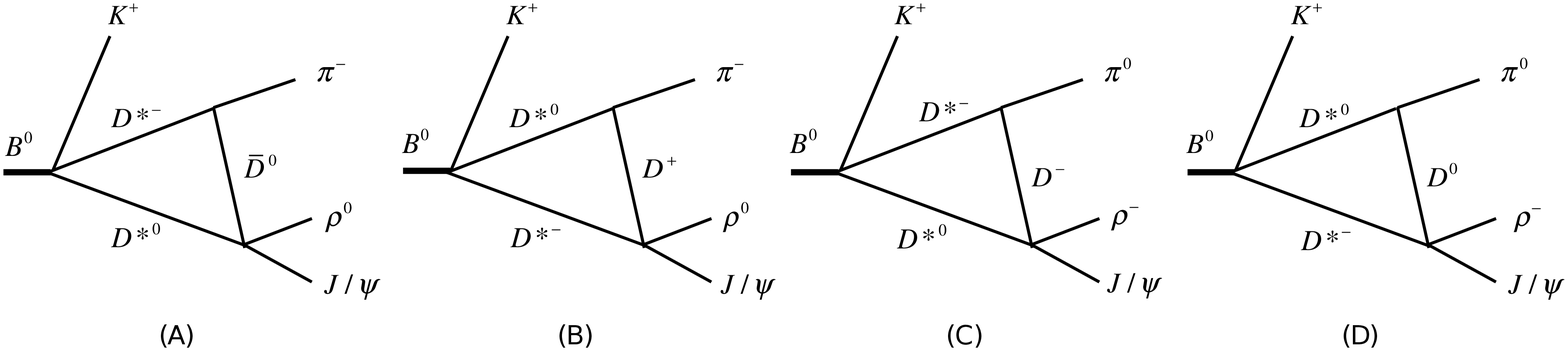}
\end{center}
 \caption{Triangle mechanisms (A,B) [(C,D)] for
 $B^{0}\to J/\psi \rho^0 K^+\pi^-$
 [$B^{0}\to J/\psi \rho^- K^+\pi^0$].
Triangle singularities from the diagrams (A,C,D) 
generate sharp peaks in
 $J/\psi \rho$ invariant mass ($M_{J/\psi \rho}$) distributions
 at $M_{J/\psi \rho}\sim 3.872, 3.876$ and 3.875~GeV, respectively.
In the text, the diagrams are referred to, 
from the left to right, 
as diagrams~A, B, C, and D, respectively.
Figures taken from Ref.~\cite{sxn_x}. Copyright (2020) APS.
 }
\label{fig:diag}
\end{figure*}

$X(3872)$~\cite{belle_x3872_jpsi-rho} is a prominent candidate of exotic hadrons,
and its internal structure has been a controversial issue.
The proposed ideas are such as
a $D^{*0}\bar{D}^0$ molecule, 
diquark-antidiquark compact tetraquark, and 
an admixture the molecule with an excited charmonium.
For getting closer to the correct understanding, 
a detailed analysis of various data relevant to 
$X(3872)$ would be important.
In this context, an issue is whether
a $X(3872)$ peak could be partly faked by a kinematical
effect called the triangle singularity (TS).
A triangle diagram like Fig.~\ref{fig:diag} 
may cause a TS
at a classical-process-like condition where 
particles in the loop are all on-shell
and they have collinear momenta with each other.
The triangle amplitude for Fig.~\ref{fig:diag}
is significantly enhanced by the TS, resulting in 
a bump in $J/\psi \rho$ and $J/\psi \rho\pi$
invariant mass distributions.
The TS has been exploited to
interpret spectrum bumps that may be due to
$XYZ$ exotic hadrons~\cite{ts_review,ts_zc4430,ts_z4050}.

For $B^0\to J/\psi \rho^0 K^+ \pi^-$,
the triangle diagram shown in Fig.~\ref{fig:diag}(A)~\footnote{
The triangle diagrams of Figs.~\ref{fig:diag}(A),
\ref{fig:diag}(B),
\ref{fig:diag}(C), and \ref{fig:diag}(D) will be refereed to as
diagrams~A, B, C, and D, respectively.}
meets the condition of a TS, and 
an $X(3872)$-like peak is expected
in the 
$M_{J/\psi \rho^0}$ distribution
($M_{J/\psi \rho^0}$: $J/\psi \rho^0$ invariant mass)
at $W\sim m_{D^{*-}}+m_{D^{*0}}$;
$W$ is the $J/\psi \rho^0\pi^-$ invariant mass and 
$m_x$ refers to the mass of a particle $x$.
The Belle experiment observed an $X(3872)$ peak
in the $M_{J/\psi\pi^+\pi^-}$ distribution of 
$B^0\to (J/\psi\pi^+\pi^-) K^+\pi^-$~\cite{belle_x3872kpi}.
In this work~\cite{sxn_x}, it is shown that 
TS from the diagram~A creates an exactly $X(3872)$-like peak.
Without adjusting any of the model parameters,
this TS peak reproduces 
the experimentally measured 
$X(3872)$ mass of $\sim$0.01\% precision and
the tightly constrained width.

This finding means that
we should take account of the TS contribution
when extracting $X(3872)$ signal from
$B^0\to J/\psi \rho^0 K^+ \pi^-$ or similar 
data in the TS region.
According to a recent proposal~\cite{guo_x3872},
the extracted $X(3872)$ signal 
accompanied by $X(3872)\pi$ and $X(3872)\gamma$ lineshapes
can be analyzed to determine the $X(3872)$ mass.
Because the non-resonant contribution from the diagram~A
is isospin conserving and not suppressed, 
an $X(3872)$-like contribution from this might be comparable to
$X(3872)\to J/\psi\pi^+\pi^-$ which is suppressed by the isospin violation.
For understanding the non-resonant mechanism and 
separating it from the $X(3872)$-pole contribution,
we also study a charge analogous 
$B^0\to (J/\psi\pi^0\pi^-) K^+ \pi^0$ decay 
where triangle diagrams C and D create an $X^-(3876)$-like peak
due to TS.

\section{Model}
The amplitude of the triangle diagram~A
for $B^0\to J/\psi\rho^0 K^+ \pi^-$ is given by
\begin{eqnarray}
 T
  &=& \int d\bm{q}\, \
  { v_{J/\psi\rho^0;D^{*0}\bar{D}^0}\, \ 
  \over
  W - E_{D^{*0}} - E_{\bar{D}^{0}} - E_{\pi^-}
  }
\Gamma_{\bar{D}^0\pi^-,D^{*-}}
{ 
 V_{K^+D^{*-} D^{*0},B^0}
\over
  W - E_{D^{*-}} - E_{D^{*0}} }
  \, ,
  \label{eq:amp}
\end{eqnarray}
where $\bm{q}$ is a loop momentum.
The energy of a particle $x$ is 
$E_x$ which depends on the particle mass ($m_x$), momentum ($\bm{p}_x$) 
and decay width ($\Gamma_x$) as 
$E_x=\sqrt{\bm{p}^2_x+m^2_x} - i\Gamma_x/2$;
$\Gamma_x$ is nonzero for $D^*$.
Similar amplitudes are given for triangle diagrams B, C, and D.
The $B^0\to J/\psi\rho^0 K^+ \pi^-$ ($B^0\to J/\psi \rho^- K^+ \pi^0$)
decay amplitude includes
the triangle mechanisms A and B (C and D) as a coherent sum.
Due to the charge-parity invariance and isospin symmetry of the strong interaction, 
the triangle mechanisms A and B (C and D)
exactly cancel each other 
if the mass and width are the same for
the charged and neutral $D^{(*)}$ mesons.
In the realistic situation, 
the TS peaks are hardly canceled
while other 
contributions outside the TS region
are significantly canceled.

The triangle diagrams~A, C, and D
cause TSs while the diagram~B does not.
This difference is caused by the fact that
$D^{*0}\to D^+\pi^-$ at on-shell is forbidden
and therefore the diagram~B does not meet 
the kinematical condition for TS.
Thus, in order to discuss TS from these triangle diagrams,
it is essentially important to take account of 
mass differences between the isospin partners
such as
$({\pi^\pm},{\pi^0})$, $({D^+},{D^0})$, and $({D^{*+}},{D^{*0}})$.
In the zero-width limit,
the TSs occur from 
the triangle amplitudes for the diagrams~A, C, and D
in the kinematical region of
$0  < M_{J/\psi\rho} - (m_{a}+m_{b}) \leq 0.2~{\rm MeV}$
and 
$0 < W - (m_{D^{*-}}+m_{D^{*0}}) \ltap 1.0~{\rm MeV}$;
$\{a,b\}=\{D^{*0},\bar{D}^{0}$\}, \{$D^{*0},D^{-}$\}, and \{$D^{*-},D^{0}$\}
for the diagrams~A, C, and D, respectively.
While TSs can be relaxed by finite widths in general, 
here the $D^{*-}$ and $D^{*0}$ widths are very small.
Thus a very sharp TS peak is expected from the diagram~A
at 
$M_{J/\psi\pi^+\pi^-}\sim m_{D^{*0}}+m_{D^{0}}=3871.7$~MeV that looks
very similar to $X(3872)$.
We also expect that
the diagrams C and D are coherently summed 
to give a sharp $X^-(3876)$-like peak in the
$M_{J/\psi\pi^0\pi^-}$ distribution.

The $D^{*\pm}$ decay width has been measured to be 
$\Gamma_{D^{*\pm}}=83.4\pm 1.8$~keV~\cite{pdg}.
Meanwhile,
only an upper limit is known for
the $D^{*0}$ decay width:
$\Gamma_{D^{*0}}<2.1$~MeV~\cite{pdg}.
$\Gamma_{D^{*0}}$ can be also calculated by 
assuming the isospin symmetry between 
$D^{*+} \to D^+\pi^0$ and $D^{*0} \to D^0\pi^0$,
and also utilizing the branching ratio of 
$D^{*0} \to D^0\gamma$ measured experimentally.
This gives $\Gamma_{D^{*0}}=55$~keV which will be adopted in our
calculations.

Regarding the interaction vertices included in Eq.~(\ref{eq:amp}),
$v_{J/\psi\rho^0;D^{*0}\bar{D}^0}$ is
an $s$-wave $D^{*0}\bar{D}^0\to J/\psi\rho^0$ interaction
from which a $J/\psi\rho^0$ pair comes out with the spin-parity
$J^P=1^+$, the spin-parity of $X(3872)$.
An $X(3872)$-pole contribution is not included in $v_{J/\psi\rho^0;D^{*0}\bar{D}^0}$.
The vertex denoted by $\Gamma_{\bar{D}^0\pi^-,D^{*-}}$ is for 
$D^{*-}\to \bar{D}^0\pi^-$.
The $B^0\to D^{*-} D^{*0} K^+$ initial decay is described by
 $V_{K^+D^{*-} D^{*0},B^0}$.
While $V_{K^+D^{*-} D^{*0},B^0}$ has a $W$-dependence,
experimental information is not available.
Because other processes such as $e^+e^-\to J/\psi D^*\bar{D}^*$
show significant enhancements near the $D^*\bar{D}^*$ threshold,
we assume that 
$V_{K^+D^{*-} D^{*0},B^0}$ has a similar $W$-dependence.
Each of the vertices include a dipole form factor on the momentum with a
cutoff $\Lambda$; $\Lambda=1$~GeV is used unless otherwise stated.
Following a standard procedure as detailed in 
Appendix~B of Ref.~\cite{3pi}, 
we calculate
the double differential decay width
$d\Gamma_{B\to J/\psi\rho^0 K\pi}/dW dM_{J/\psi\rho}$
with the decay amplitude of Eq.~(\ref{eq:amp}).
We then take account of the $\rho^0\to\pi^+\pi^-$ decay, giving
$d\Gamma_{B^0\to J/\psi\pi^+\pi^-K^+\pi^-} / dW dM_{J/\psi\pi^+\pi^-}$.

\begin{figure*}[t]
\begin{center}
\includegraphics[width=1\textwidth]{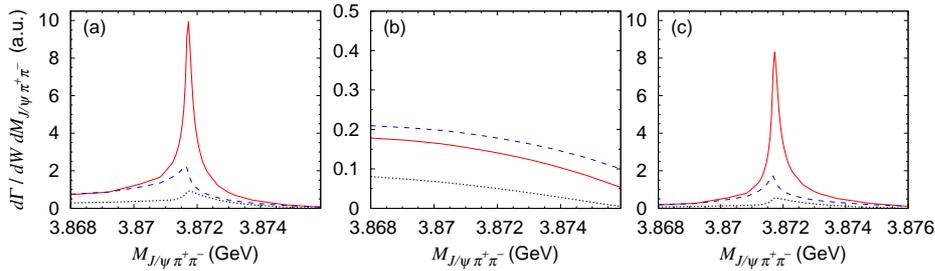}
\end{center}
\caption{
 $J/\psi\pi^+\pi^-$ invariant mass ($M_{J/\psi\pi^+\pi^-}$)
 distributions for $B^{0}\to J/\psi\pi^+\pi^-K^+\pi^-$
 given by the triangle diagrams of Figs.~\ref{fig:diag}(A)
 and \ref{fig:diag}(B).
$M_{J/\psi\pi^+\pi^-}$ is from 
 $J/\psi$ and $\pi^+\pi^-$ from $\rho^0$ decay.
The spectra at 
$W-(m_{D^{*0}}+m_{D^{*-}})=-1.0$, 0.0, and 1.2~MeV are shown by 
the black dotted, red solid, and blue dashed curves,
respectively;
 $W$ denotes the invariant mass of the final $J/\psi\pi^+\pi^-\pi^-$
 subsystem and $m_{D^{*0}}+m_{D^{*-}}=4017.1$~MeV.
The panels (a) and (b) show the spectra from
the triangle diagrams of Figs.~\ref{fig:diag}(A) and \ref{fig:diag}(B),
respectively;
the panel (c) shows their coherent sum.
Although the overall normalization of the figures is arbitrary, 
the relative heights among the
 curves in all the figures
 are a prediction of the model.
Figures taken from Ref.~\cite{sxn_x}. Copyright (2020) APS.
}
\label{fig:spec}
\end{figure*}

\section{Results}

The double differential decay width
$d\Gamma_{B^0\to J/\psi\pi^+\pi^-K^+\pi^-} / dW dM_{J/\psi\pi^+\pi^-}$
are shown 
as a function of $M_{J/\psi\pi^+\pi^-}$ 
in 
Figs.~\ref{fig:spec}(a), \ref{fig:spec}(b), and \ref{fig:spec}(c)
for the triangle diagrams~A, B, and A+B, respectively.
These spectra are for the TS region  ($W\sim m_{D^{*0}}+m_{D^{*-}}$) and around.
We find the outstanding peak 
created by the TS from the diagram~A
at $M_{J/\psi\pi^+\pi^-}\sim 3871.7$~MeV.
The peak position seems 
exactly the
precisely measured $X(3872)$ mass:
$3871.69\pm 0.17$~MeV~\cite{pdg}.
This peak position and the narrow width
is a parameter-free prediction from 
the triangle diagrams~A.
We examined 
the cutoff dependence over $\Lambda=0.5-2$~GeV, 
and confirmed the stability of the position and shape of
the TS peak.
The other arbitrary parameters
change the overall normalization only.
The TS peak is such stable because:
(i) the $D^*$ width is tiny and thus the TS is very close to the
physical region and thus a dominant effect;
(ii) details of the dynamics play little role for the TS.
Figure~\ref{fig:spec} also shows that the TS peak height changes
sensitively to $W$ because the TS occurs in the small $W$ window as
discussed above.

Figure~\ref{fig:spec}(b) shows that
the triangle diagram~B generates
smooth lineshapes at $W\sim m_{D^{*0}}+m_{D^{*-}}$.
The diagrams~A and B look similar but they behave very differently. 
This difference is due to 
the fact that only the diagram~A meets 
the TS condition.
The tiny $D^*$ width is also a reason for this 
high selectivity of the TS condition.
After taking the coherent sum of 
the diagrams~A and B, as shown in Fig.~\ref{fig:spec}(c),
the diagram B cancel the smooth background-like contribution from 
the diagram~A.

\begin{figure}[t]
\begin{center}
\includegraphics[width=.8\textwidth]{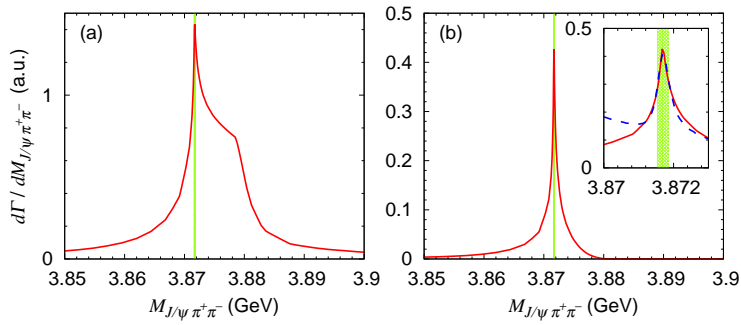}
\end{center}
 \caption{
$M_{J/\psi\pi^+\pi^-}$
 distributions given by the coherently summed triangle diagrams of Figs.~\ref{fig:diag}(A)
 and \ref{fig:diag}(B).
(a) The red solid curve is obtained by 
integrating the spectra in 
 Fig.~\ref{fig:spec}(c) and those in higher $W$ region 
with respect to $W$ over the whole $W$-region.
(b) The red solid curve is obtained by the $W$-integral in the range of
$-3~{\rm MeV} \le W - (m_{D^{*-}}+m_{D^{*0}}) \le 4~{\rm MeV}$.
The peak region is shown in the insert.
 The blue dashed curve shows the Breit-Wigner plus a background fitted to the red solid curve.
The $X(3872)$ mass range from the PDG~\cite{pdg} 
is indicated by the green bands.
Figures taken from Ref.~\cite{sxn_x}. Copyright (2020) APS.
 }
\label{fig:peak}
\end{figure}

The Belle data for $B^{0}\to J/\psi\pi^+\pi^-K^+\pi^-$~\cite{belle_x3872kpi}
shows a peak at $M_{J/\psi\pi^+\pi^-}\sim 3.872$~GeV
in $d\Gamma_{B^0\to J/\psi\pi^+\pi^-K^+\pi^-} / dM_{J/\psi\pi^+\pi^-}$.
This data includes the whole $W$ region allowed kinematically.
We calculate a theoretical counterpart
by integrating the spectra in 
Fig.~\ref{fig:spec}(c) and also those in higher $W$ region with respect to $W$.
Figure~\ref{fig:peak}(a) is the obtained spectrum.
While a sharp peak remains at $M_{J/\psi\pi^+\pi^-}\sim 3.872$~GeV,
there is 
also a large shoulder near the $D^{*-}D^+$ threshold.
This shoulder is from the threshold cusp generated by the diagram~B.
This spectrum is smeared with the experimental resolution and compared
with the Belle data.
The lineshape is found to be too broad to explain the data.
Although 
this lineshape depends on the $W$-dependence assumed for the
$B^0\to D^{*-} D^{*0} K^+$ vertex,
the diagrams~A and B
does not seem likely to
explain the Belle data.

Now the $W$-integral is limited to the TS region and around:
$-3~{\rm MeV} \le W - (m_{D^{*-}}+m_{D^{*0}}) \le 4~{\rm MeV}$.
The resulting spectrum is given by
the red solid curve in Fig.~\ref{fig:peak}(b).
A single  narrow peak clearly appears.
To quantify the peak position and width, we fit the spectrum
with 
the common resonance($X$)-excitation mechanism,
$B\to X K\pi$ and $X \to J/\psi\rho^0$.
The fit gives the Breit-Wigner mass and width for $X$.
A coherent background is also added with
an adjustable quadratic polynomial of $M_{J/\psi\pi^+\pi^-}$.
The fit result is given by the blue dashed curve in
the small insert of Fig.~\ref{fig:peak}(b).
The fit does not work well in the tail region
because: 
(i) the Breit-Wigner form and the spectrum shape are rather different;
(ii) the background rapidly decrease near the higher end of
the $M_{J/\psi\pi^+\pi^-}$ distribution
due to the available phase-space.
Nevertheless, 
the obtained Breit-Wigner mass and width still quantify
the peak, and are
 $m_{\rm BW}=3871.68\pm 0.00$~MeV and 
$\Gamma_{\rm BW}=0.42\pm 0.01$~MeV, respectively;
the ranges of the values are from the cutoff dependence.
The Breit-Wigner mass value is larger than the $D^{*0}\bar{D}^0$ threshold
by only a few keV.
The Breit-Wigner parameters can be compared
with the PDG value for $X(3872)$:
 $m_{\rm PDG}=3871.69\pm 0.17$~MeV and $\Gamma_{\rm PDG}< 1.2$~MeV.
These Breit-Wigner parameters have
a very small cutoff dependence, and 
agree very well 
with the precise experimental measurement for $X(3872)$.
From the above results, it is indicated that
the TS peak 
from the diagram~A might partly fake the $X(3872)$ signal
in $B^{0}\to J/\psi\pi^+\pi^-K^+\pi^-$
at $W\sim m_{D^{*-}}+m_{D^{*0}}$.
Thus, when applying the $X(3872)$ mass determination
method of Ref.~\cite{guo_x3872} to data, 
one needs to consider a possible contribution from the non-$X$ triangle
mechanisms.

\begin{figure*}[t]
\begin{center}
\includegraphics[width=.8\textwidth]{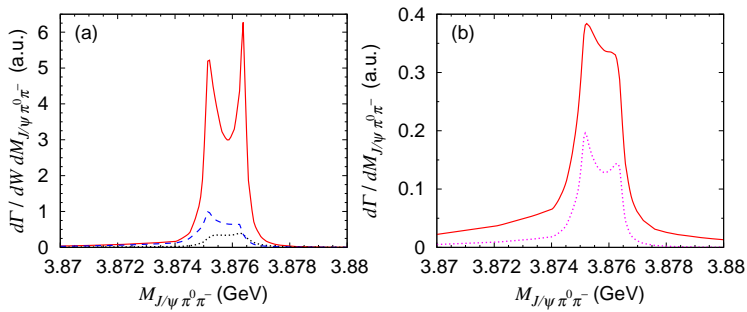}
\end{center}
\caption{
$M_{J/\psi\pi^0\pi^-}$ distributions for $B^{0}\to J/\psi\pi^0\pi^-K^+\pi^0$.
The triangle diagrams of
Figs.~\ref{fig:diag}(C) and \ref{fig:diag}(D)
are coherently summed to generate the spectra.
$M_{J/\psi\pi^0\pi^-}$ is from 
 $J/\psi$ paired with $\pi^0\pi^-$ from $\rho^-$ decay.
(a) $M_{J/\psi\pi^0\pi^-}$ distributions for 
$W-(m_{D^{*0}}+m_{D^{*-}})=-1.0$, 0.0, and 1.2~MeV
are shown by 
the black dotted, red solid, and blue dashed curves, respectively.
 (b) The spectra
 in the panel (a) are integrated with respect to $W$ to give 
the red solid curve.
The integral in the limited region of $-3~{\rm MeV} \le W - (m_{D^{*-}}+m_{D^{*0}}) \le 4~{\rm MeV}$
gives the magenta dotted curve.
The overall normalization of 
the curves in the panel (a) is the same as 
those in Fig.~\ref{fig:spec}.
Figures taken from Ref.~\cite{sxn_x}. Copyright (2020) APS.
}
\label{fig:spec5}
\end{figure*}
A charge analogous process, 
$B^0\to (J/\psi\pi^0\pi^-)K^+\pi^0$,
from the triangle diagrams C and D in Fig.~\ref{fig:diag} is now considered.
Around the TS region ($W\sim m_{D^{*-}}+m_{D^{*0}}$),
the triangle mechanisms generate
the $M_{J/\psi\pi^0\pi^-}$ distribution as 
in Fig.~\ref{fig:spec5}(a).
Each of the triangle diagrams C and D hits a TS to generate a sharp peak.
Their coherent sum appears as 
the clear twin peaks in the figure.
We see again the acute $W$ dependence of the spectra.
We integrate the spectra
in Fig.~\ref{fig:spec5}(a) and those in higher $W$ region 
with respect to $W$, 
the red solid curve in Fig.~\ref{fig:spec5}(b) is obtained.
When we limit the $W$-integral near the TS region only, 
the magenta dotted curve is obtained.
We find the clear peak after the $W$-integral.
A future experiment could find this $X^-(3876)$-like 
peak.
If the peak is observed experimentally, this would be a clear
identification of TS in data because 
no charged and very narrow resonance is known in this energy region.

The non-resonant $X(3872)$-like peak is generated by the diagram~A
in the TS region. 
When $X(3872)$ signal is analyzed by the method of Ref.~\cite{guo_x3872}
to determine the $X(3872)$ mass,
this TS contribution has to be eliminated in advance. 
We proposed a method to estimate the non-resonant contribution by
studying the charge analogous process. Details are discussed in Ref.~\cite{sxn_x}.



\end{document}